\documentclass[twocolumn,superscriptaddress]{revtex4}%%,twocolumn
\usepackage{graphicx}
\usepackage{times}
\usepackage{amsmath}
\usepackage{amssymb}
\usepackage{units}
\usepackage{stmaryrd} %% "// sign" = \sslash
\usepackage[compact]{titlesec}
\titlespacing{\section}{0pt}{*0}{*0}
\titlespacing{\subsection}{0pt}{*0}{*0}
\titlespacing{\subsubsection}{0pt}{*0}{*0}

\begin{document}
\preprint{0}

\title{2D surprises at the surface of 3D materials: confined electron systems in transition metal oxides}

\author{Emmanouil Frantzeskakis}
\thanks{emmanouil.frantzeskakis@csnsm.in2p3.fr}
\address{CSNSM, Universit\'e Paris-Sud, CNRS/IN2P3, Universit\'e Paris-Saclay, 91405 Orsay cedex, France}

\author{Tobias Chris R\"odel}
\address{CSNSM, Universit\'e Paris-Sud, CNRS/IN2P3, Universit\'e Paris-Saclay, 91405 Orsay cedex, France}
\address{Synchrotron SOLEIL, L'Orme des Merisiers, Saint-Aubin - BP 48, 91192 Gif-sur-Yvette, France}

\author{Franck Fortuna}
\address{CSNSM, Universit\'e Paris-Sud, CNRS/IN2P3, Universit\'e Paris-Saclay, 91405 Orsay cedex, France}

\author{Andr\'es Felipe Santander-Syro}
\thanks{andres.santander@csnsm.in2p3.fr}
\address{CSNSM, Universit\'e Paris-Sud, CNRS/IN2P3, Universit\'e Paris-Saclay, 91405 Orsay cedex, France}

%\date{\today}

\begin{abstract}
The scope of this article is to review the state-of-the-art in the field of confined electron systems generated at the bare surfaces of transition metal oxides (TMOs). This scientific field is a prime example of a domain where two-dimensional physics and photoemission-based spectroscopic techniques have together set up the development of the story. The discovery of a high-mobility two-dimensional electron system (2DES) at interfaces of transition metal oxides has attracted an immense scientific interest due to new opportunities opened in the emerging field of oxide electronics. The subsequent paradigm shift from interfaces to the bare surfaces of TMOs made the confined electron system accessible to surface-sensitive spectroscopic techniques and this new era is the focus of the present article. We describe how results by means of Angle-Resolved Photoemission Spectroscopy (ARPES) establish the presence of confined electron carriers at the bare surface of SrTiO$_{3}$(100), which exhibit complex physics phenomena such as orbital ordering, electron-phonon interactions and spin splitting. The key element behind the 2DES generation is oxygen vacancies. Moreover, we review the experimental evidence on the generation of 2DESs on surfaces with different orientation, as well as on different TMO substrates. The electronic structure of the confined electron system responds to such changes, thereby providing external means for engineering its properties. Finally, we identify new directions for future research by introducing a device-friendly fabrication protocol for the generation of 2DESs on TMO surfaces.
\end{abstract}

\maketitle

\section{Introduction}

In scientific literature, the term ``functional oxides'' traditionally refers to the family of transition metal oxides (TMOs) composed of oxygen anions and metals with partially-filled d-shells. TMOs are of high technological importance as catalysts \cite{Kung1989}, superconductors \cite{Bednorz1986, Damascelli2003}, novel platforms for electronics \cite{Mannhart2010} and building blocks for spintronic devices \cite{Salamon2001}. Moreover, in terms of accessing novel physics phenomena, the interplay of $d$ electrons in transition metal oxides offers access to diverse quantum ground states including metallic and semiconducting phases, (anti-)ferromagnetism, ferroelectricity, multiferroic behavior, superconductivity, metal-insulator transitions, charge ordering and orbital ordering \cite{Cheong2007}.

The interest of the scientific community on transition metal oxides was rekindled in 2004 by the intriguing possibility to exploit certain oxide interfaces as platforms for oxide electronics \cite{Mannhart2010, Takagi2010}. The discovery of a high mobility two-dimensional electron system (2DES) on the interface between two otherwise insulating oxides, namely LaAlO$_3$ (LAO) and SrTiO$_3$ (STO) \cite{Ohtomo2004}, drove the subsequent scientific effort towards exploiting the unexpected 2DES and understanding its emergence \cite{Thiel2006, Caviglia2008, Nakagawa2006, Kalabukhov2007, Slooten2013}. The 2DES at the LAO/STO interface is very important not only because it gives rise to 2D conductivity but also because it is at the origin of magnetism, superconductivity and their coexistence, all of them controllable by a gate-voltage \cite{Reyren2007, Brinkman2007, Li2011, Bert2011}. This system is therefore a vivid illustration of exotic physics phenomena in low dimensions and has a high potential for technological applications. In short, LAO/STO has been considered as the prime candidate for oxide electronics not only because it is the most well-studied oxide system which hosts a high-mobility 2DES, but also due to the scarcity of alternative interfaces \cite{Perna2010, Jang2011, Hotta2007}. Its high potential led to the term ``functional oxides'' being used primarily for variations of the LAO/STO interface.

A second set of breakthrough studies came in early 2011 and showed that the LAO/STO interface is not necessary to obtain a high mobility 2DES: a similar 2DES was discovered at the bare (100) surface of SrTiO$_3$ \cite{Santander2011, Meevasana2011}. These results opened a whole new line of research. The 2DES became accessible to surface sensitive techniques as it was no longer buried under layers of LaAlO$_3$ but it could be probed right on the sample surface -and subsurface layers. This is in contrast to interface systems where access to the confined electronic structure is very difficult and with poor resolution \cite{Berner2013, Plumb2013, Cancellieri2016}. For 2DESs at TMO surfaces, Angle Resolved Photoemission Spectroscopy (ARPES) became the leading experimental technique in this field due to its high surface sensitivity and the possibility for a direct and high-resolution view of the electronic structure of the 2DES. In the present work we will show how this spectroscopic technique resulted in a deep understanding of the confined electron systems on the TMO surfaces and paved the way for further research. Our goal is to present a combined overview of ARPES studies in this field and bring into focus the latest advances made by our research team at CSNSM. We will start by discussing the characteristics of the 2DES on SrTiO$_3$(100): details of its electronic structure can lead to an understanding of the 2DES origin and reveal the effect of more advanced concepts such as many-body interactions. After reviewing the basic and more advanced characteristics of the 2DES on SrTiO$_3$(100), we will discuss 2DESs on other TMO surfaces: different surfaces of STO give the possibility of orientational tuning of the 2DES, KTaO$_{3}$ reveals the effect of bulk spin-orbit coupling, TiO$_{2}$ anatase is an example of a non-perovskite TMO and BaTiO$_{3}$ shows that a ferroelectric compound can also host a 2DES. Finally, we will finish by bringing into focus a new versatile technique with a high potential of generating device-friendly TMO/2DES systems.\\

\section{Experimental Details}

This section is a summary of the experimental details on sample preparation and on ARPES measurements performed by our team at CSNSM. Further details are explained in Ref. \onlinecite{Rodel_phd} or in the references mentioned in the figure captions.

Transition metal oxides went through various steps of \textit{ex-situ} and \textit{in-situ} treatment. Before introducing them into the UHV setup, SrTiO$_{3}$ single crystals were etched in buffered HF to create large TiO$_{2}$-terminated terraces and they were subsequently annealed under O$_{2}$ flow to heal the etch holes with a good control of bulk oxygen vacancies. In order to obtain a clean and well-ordered surface, \textit{in-situ} treatment of the different TMO surfaces requires either fracturing or a combination of ion bombardment and/or annealing in UHV conditions. The detailed protocol of \textit{in-situ} preparation varies for different TMOs. SrTiO$_{3}$ was prepared either by fracturing or annealing. For the latter we used temperatures of the order of 600$^{\textmd{o}}$C for 1-2 h as higher temperatures or longer annealing times are known to promote surface reconstructions \cite{Naito1994, Frantzeskakis2012, Chang2008}. Clean surfaces of KTaO$_{3}$ and TiO$_{2}$ anatase were exclusively produced by fracturing in UHV conditions. In all cases, the cleanliness and order of the surfaces were checked by Auger Electron Spectroscopy (AES) and Low Energy Electron Diffraction (LEED). Monodomain BaTiO$_{3}$ thin films were grown along the $<$001$>$ direction by Pulsed Laser Deposition (PLD) using a Kr-F excimer laser. The growth of the film was monitored by RHEED. Prior to ARPES experiments the films were annealed at 550$^{\textmd{o}}$C to remove contamination produced during sample transfer. 

After obtaining a clean surface, additional capping with pure aluminum was used in the cases of SrTiO$_{3}$, anatase and BaTiO$_{3}$ in order to generate oxygen vacancies through a redox reaction (see part E of the next section). Aluminum was depositied by means of a Knudsen cell with an alumina crucible. The approximate growth rate was 0.3 \AA/min and the flux was calibrated by means of a quartz microbalance. In order to create a high concentration of oxygen vacancies on the near-surface region, amorphous Al films with a thickness of 2 $\textmd{\AA}$ were deposited on the clean TMO surfaces.
 
ARPES experiments were performed at the Synchrotron SOLEIL (France) and at the Synchrotron Radiation Center (SRC, University of Wisconsin, Madison) using Scienta R4000 hemispherical analyzers with vertical slits. We used variable photon energy, typically in the range of 20-120 eV both with linear horizontal and linear vertical polarization. Measurements were performed at 8K (SOLEIL) or at 25K (SRC) without observing any major T-dependence between these two temperature values. In all cases the pressure was in the range of 10$^{-11}$ mbar. The typical angle and energy resolutions were 0.25$^{\textmd{o}}$ and 15 meV, respectively, while the mean diameter of the incident photon beam was of the order of 50 $\mu$m (SOLEIL) and 150 $\mu$m (SRC).\\\\\\\

\section{Results and Discussion}
\subsection{2DES on a bare surface of SrTiO$_{3}$: discovery and basic characteristics}

\begin{figure*}
  \centering
  \includegraphics[width = 13 cm]{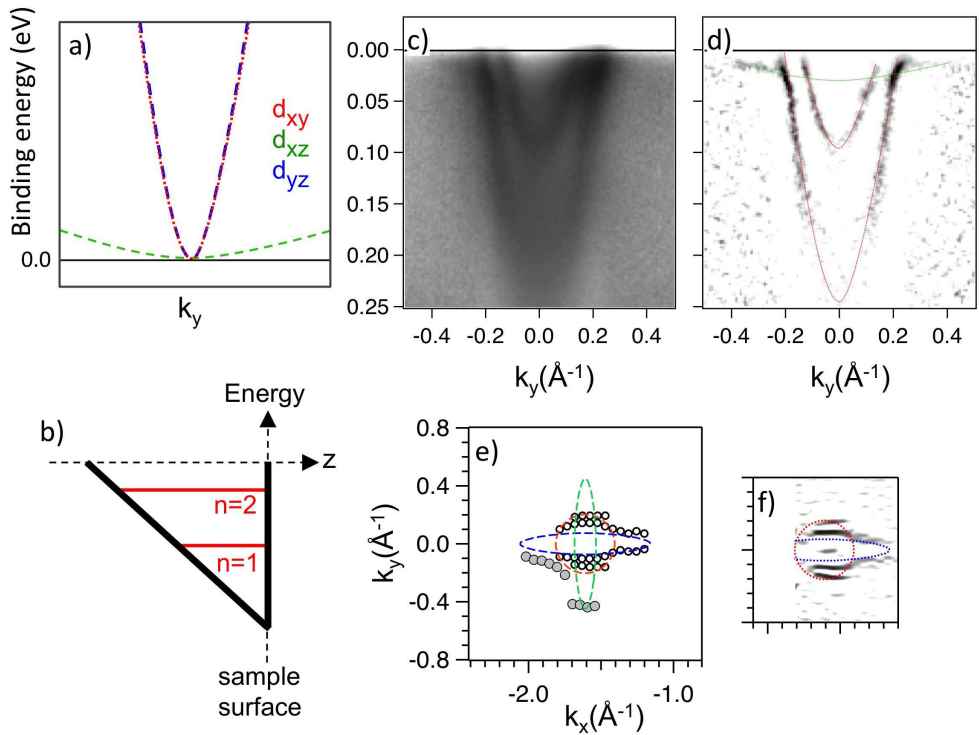}
  \caption{(a) The bulk electronic structure of SrTiO$_{3}$ along $k_y$. $d_{xy}$, $d_{xz}$ and $d_{yz}$ orbitals give rise to a heavy band along the $k_{z}$, $k_{y}$ and $k_{x}$ direction, respectively. In the absence of spin-orbit coupling, two light and one heavy band are degenerate at the center of the Brillouin zone (i.e. $k_y=0$). The bands lie above the Fermi level and they are hence unoccupied. (b) A depth-dependent potential acts as a quantum well that splits each state into a series of subbands. The actual potential is modeled by a triangular quantum well where $z=0$ marks the position of the sample surface. (c) Experimental ARPES data showing the effect of the depth-dependent potential on the surface electronic structure of SrTiO$_{3}$. In contrast to the bulk electronic structure [panel (a)], the bands are shifted below the Fermi level and two $d_{xy}$ subbands are clearly visible. (d) Same as panel (c) using the 2D curvature of the experimental data \cite{Zhang2011}. The heavy $d_{xz}$ band is also visible. (e) Experimental Fermi surface of SrTiO$_{3}$(100) consisting of one circular and two elliptical contours that are orthogonal two each other. Dashed lines are guides to the eye and track the Fermi momenta (white and grey circles) acquired by high-statistics measurements at the center of various Brillouin zones. Colors correspond to the orbital origin of the contours following panel (a). (f) Experimental Fermi surface of SrTiO$_{3}$(100) by plotting the 2$^{\textmd{nd}}$ derivative of the ARPES data. Data have been acquired using 47 eV photons and linear vertical polarization. Part of the figure has been adapted from Ref. \onlinecite{Santander2011}.}
\end{figure*}

An experimental ARPES work by the CSNSM team and collaborators \cite{Santander2011}, along with a contemporary study by Meevasana et al. \cite{Meevasana2011} were the first studies to report a 2DES on a bare TMO surface setting the scientific background for further work. The key idea behind these studies is that an electric field at the sample surface lifts the degeneracy of the $t_{2g}$ states and confines them into a series of subbands by generating a depth-dependent potential that acts as a quantum well [Figs. 1(a)-1(d)]. Santander et al. approximated the confining potential by a potential wedge $V_{\textmd{conf}}=(z-d)F$, where $F$ is the strength of the electric field, $z=0$ corresponds to the sample surface and $z=d$ corresponds to the maximum depth of $V_{\textmd{conf}}(z)$ \cite{Santander2011}. Under this approximation, the solution of the 1D Hamiltonian for non-interacting electrons trapped in this potential well yields the quantized eigenenergies of the confined subbands \cite{Santander2011}:
\begin{equation}
E_{n}=V_{\textmd{conf}}(0)+\left(\frac{\hbar^{2}}{2m_{z}^{*}}\right)^{-\frac{1}{3}}\left[\left(\frac{3\pi}{2}\right)\left(n-\frac{1}{4}\right)eF\right]^{\frac{2}{3}}
\end{equation}

where $n$ is the index of the quantized subbands, $e$ is the electron charge, $m_{z}^{*}$ is the effective mass along the out-of-plane direction.

From the experimental energy splitting of two successive subbands (e.g. $E_2$$-$$E_1$), Santander et al. could determine the electric field strength $F=83$ MV/m and consequently the bottom of the confining potential $V_{\textmd{conf}}(0)=-260$ meV, as well as its spatial extension $d=31$$\textmd{\AA}$ \cite{Santander2011}. In the same work, the value of the confining field strength was independently cross-checked by using the Gauss theorem and the carrier density directly measured from the area enclosed by the Fermi surface of the 2DES. Going beyond the simplified but pedagogical wedge approximation, Meevasana et al. performed Poisson-Schr\"odinger calculations to estimate the band bending profile that acts as a confining potential \cite{Meevasana2011}. The best fit to the experimental data was obtained for $V_{\textmd{conf}}(0)=-700$ meV where the lowest subband is localized within $\sim20$$\textmd{\AA}$ from the surface and it is followed by a series of bands with smaller binding energy and deeper extension into the bulk. The calculated band bending profile is rather steep as expected for a system with strongly confined two-dimensional states. A band bending profile that extends further into the bulk -as it would be expected for a more three-dimensional system- fails dramatically to simulate the experimental spectra. The two-dimensional character of the observed states can be therefore inferred by the agreement of the experimental ARPES data with the predicted energy dispersion of confined states in a steep electrostatic potential. 

There are further arguments in favor of the low dimensionality of the observed states. First of all, the area of the experimentally determined Fermi surface consisting of $d_{xy}$, $d_{xz}$ and $d_{yz}$ orbitals [Figs. 1(e), 1(f)] yields a surface carrier density $n_{\textmd{2D}}$ of the order of 10$^{14}$ cm$^{-2}$. If these carriers were three-dimensional, the corresponding carrier density $n_{\textmd{3D}}$ would be of the order of 10$^{21}$ cm$^{-3}$ a value which is much higher than the bulk doping of the samples used in Ref. \onlinecite{Santander2011}. As a matter of fact, the experimental Fermi surface is insensitive to the value of bulk doping, while the latter varies between 10$^{13}$ cm$^{-3}$ and 10$^{21}$ cm$^{-3}$ \cite{Santander2011}, thereby pointing towards a surface origin. Moreover, despite their strong dispersion in the $k_{x}$-$k_{y}$ plane, both studies agree that the observed electronic states do not exhibit any appreciable dispersion along the surface normal, as expected for states with 2D character \cite{Santander2011, Meevasana2011}. From the above, it is evident that the (100) surface of SrTiO$_{3}$ hosts charge carriers that are not related to the bulk. The presence of a 2DES with high carrier density is therefore well established.

\begin{figure}
  \centering
  \includegraphics[width = 7 cm]{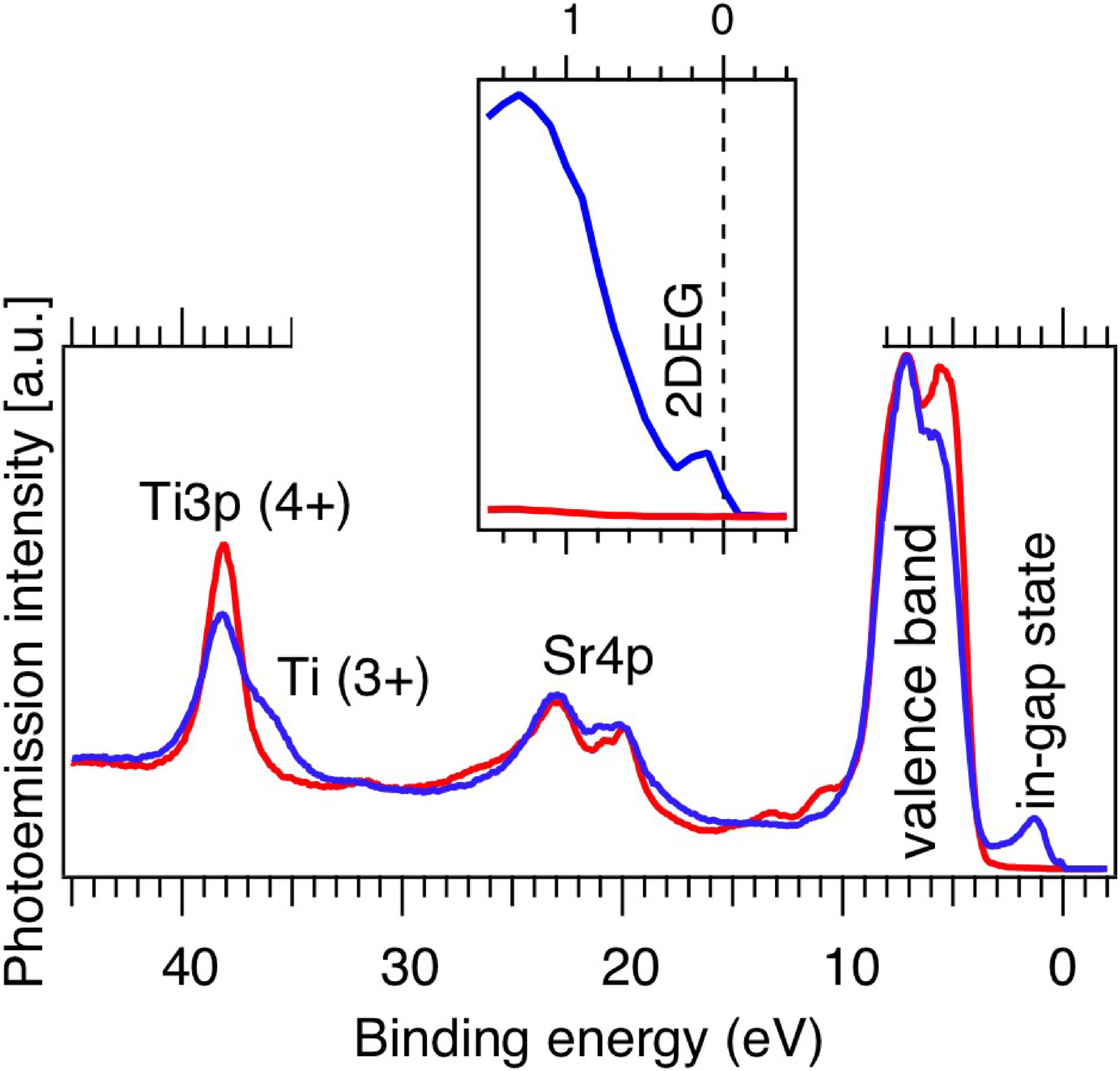}
  \caption{Angle integrated photoemission spectra acquired on a SrTiO$_{3}$ surface before (red) and after (blue) long exposure to ultraviolet synchrotron radiation. A 2DES (blue spectrum inset) is accompanied by the generation of an in-gap state related to oxygen vacancies and by spectral changes in the Ti$3p$ peak and the valence band. All SrTiO$_{3}$ surfaces present the same changes upon UV exposure: these data have been acquired on a (111)-oriented surface of SrTiO$_{3}$.}
\end{figure}

We now turn to the origin of the confining potential and the 2DES charge carriers. The two most widely accepted models for the origin of a 2DES on TMO interfaces are the creation of oxygen vacancies \cite{Kalabukhov2007, Slooten2013} and the electronic reconstruction to avoid an interfacial polar discontinuity (i.e. the polar catastrophe) \cite{Nakagawa2006}. The fact that confined conduction carriers appear at the bare surface of STO(100) which is non-polar cannot be reconciled with the polar catastrophe scenario. In fact, already from the first ARPES studies of a 2DES on STO(100), the role of oxygen vacancies was well established. First of all, there are clear spectral changes in the O$2p$ valence band during the development of the 2DES. An energy shift of its leading edge with respect to the value determined by bulk-sensitive optical adsorption shows that the O$2p$ states are subject to surface band bending \cite{Santander2011}. Moreover, gradual loss of the O$2p$ spectral weight may be a hint of oxygen vacancy creation at the surface \cite{Meevasana2011}. Loss of spectral weight is observed at the low binding energy ($E_{\textmd{b}}$) side of the O$2p$ feature that consists of states which are most likely to change their original environment: O$_{\pi\pi}$ states of non-bonding character or with negligible hybridization with $d$ orbitals \cite{Asahi2000}. Changes in the valence band are in line with other spectral modifications. Namely, the low-$E_{\textmd{b}}$ side of the Ti$3p$ peak develops a clear shoulder. This is a fingerprint of Ti ions having lower oxidation numbers - Ti$^{3+}$ instead of nominal Ti$^{4+}$ due to potential oxygen depletion. Moreover, the appearance of the 2DES is accompanied by another electronic state at an approximate binding energy of 1.3 eV. This state lies in the bulk electronic gap of SrTiO$_{3}$ and was attributed to oxygen vacancies at the surface \cite{Aiura2002}. Figure 2 summarizes the spectral differences between STO surfaces with and without a 2DES. In agreement with expectations from an oxygen-vacancy scenario, Aiura et al. had already shown in 2002 that when the STO(100) surface is exposed to oxygen, the electronic state at the Fermi level is suppressed, the O$2p$ valence band shifts back to lower binding energies and the in-gap state disappears \cite{Aiura2002}. From the above, it is evident that a high concentration of positively charged oxygen vacancies in the near-surface region is at the origin of the electrostatic potential that bends the conduction band below the Fermi energy ($E_{\textmd{F}}$) and confines its electrons. As a matter of fact, the electronic structure of the $3d$ orbitals appears very differently in the bulk and at the surface. In the bulk the $t_{2g}$ conduction band is unoccupied and three-dimensional, while on the surface it becomes partially occupied and yields two-dimensional carriers. We finally note that various DFT calculations have confirmed the importance of oxygen vacancies in inducing a 2DES at the surface of SrTiO$_{3}$ \cite{Santander2011, Shen2012, Silva2014, Zhuang2014}. 

Having established oxygen vacancies as the key element for the 2DES at the bare surface of STO(100), another question naturally arises: how are these oxygen vacancies created? Annealing in UHV during sample preparation can introduce oxygen vacancies \cite{Pal2014} but this is not the decisive phenomenon for the generation of a 2DES due to many reasons. First of all, we note that a first set of ARPES studies on STO(100) was performed on samples that were fractured in UHV rather than surfaces prepared in-situ by annealing \cite{Santander2011, Meevasana2011}. Secondly, oxygen vacancies induced by UHV annealing are distributed rather homogeneously all over the volume of the sample and they do not reside exclusively in the near-surface region \cite{Souza2012}. We note that carrier accumulation near the surface is a necessary prerequisite in order to have a surface-to-bulk electrostatic potential that gives rise to band bending and quantum confinement. Finally, an ARPES study by Plumb et al. casted away any remaining doubts on the necessity of high temperature annealing: it was shown that the observed confined states are identical for different UHV annealing \cite{Plumb2014}. Moreover, these states are identical to their counterparts observed on cleaved SrTiO$_{3}$ surfaces \cite{Santander2011, Meevasana2011} and bulk doping has no effect on the near-$E_{\textmd{F}}$ electronic structure \cite{Santander2011}. The 2DES seems therefore insensitive to the history of the sample as long as the surface is clean and well-ordered. Oxygen vacancies must be therefore created at later stages. As a matter of fact, Meevasana et al. showed that the 2DES charge density starts from low values and increases with the irradiation dose \cite{Meevasana2011}. One can naturally assume that ultraviolet light induces oxygen desorption and hence the creation of oxygen vacancies. The advantage of this method for the generation of oxygen vacancies is that one can control the charge carrier density by tuning the excitation source \cite{Meevasana2011}, albeit at the price of a non-homogeneous 2DES when the sample is not irradiated uniformly. 

Up to this point all pieces of the puzzle fall well into place for the 2DES observed on the surface of STO(100): 
\begin{itemize}
  \item[$\cdot$] incoming ultraviolet radiation creates oxygen vacancies at the surface
    \item[$\cdot$] oxygen vacancies provide a high concentration of near-surface positively charged defects and excess electrons
        \item[$\cdot$] the positively charged defects create a surface-to-bulk electrostatic field and potential 
        \item[$\cdot$] the electrostatic field/potential induces band bending
        \item[$\cdot$] the steep band bending depth profile can confine any charge carriers
         \item[$\cdot$] the formerly unoccupied $t_{2g}$ conduction band follows the profile of the electrostatic potential and bends below $E_{\textmd{F}}$ in the near-surface region, thus giving rise to conduction electrons
         \item[$\cdot$] these extra conduction electrons are due to the excess electrons from the oxygen vacancies
          \item[$\cdot$] the conduction electrons see the depth profile of the electrostatic potential as a quantum well along the surface normal and they become confined along this direction
         \item[$\cdot$] as the surface normal is the only direction of confinement, conductions electrons are free to move only along a two-dimensional plane: in other words they form a 2DES
         \item[$\cdot$] the spectroscopic fingerprint of the 2DES are subbands with appreciable energy-momentum dispersion along any direction parallel to the surface plane, but at the same time with well-determined energy values due to quantum confinement along the surface normal
\end{itemize}

The studies that signalled the discovery of a 2DES on STO(100) \cite{Santander2011, Meevasana2011} were however only the spark. Later research built on this knowledge and brought the field into a more profound understanding. We will summarize these results giving emphasis to ARPES studies that followed various research paths:\\
(i) 	the experimental observation of 2DES characteristics that remained elusive (e.g. many-body interactions, spin polarization, further details on the creation mechanism, on energy dispersion and on dimensionality).\\  
(ii) 	the investigation of STO surfaces with different orientations.\\
(iii)	the generation of 2DES on different functional oxides.\\
(iv)	the demonstration of a device-friendly fabrication protocol that may bridge the gap between ARPES studies and technological applications.  \\

\subsection{2DES at the bare surface of SrTiO$_{3}$: towards a more detailed understanding}
In order to shed more light into the 2DES, the role of irradiation in creating a metallic surface was further investigated. Plumb and coworkers observed that on increasing irradiation, the size of the experimental Fermi surface (and hence the carrier density) saturates very quickly, while intensity continues increasing. They proposed that increasing intensity signals a larger portion of the sample becoming metallic and conclude that there are two stable configurations of STO(100): one insulating (no irradiation, no 2D carriers, no Fermi surface) and one with a fixed density of metallic 2D carriers \cite{Plumb2014}. A later work by McKeown Walker et al. showed that there are not only two stable electronic configurations of STO(100) but that $n_{\textmd{2D}}$ can be continuously tuned as a function of the irradiation time (fluence) and the photon energy \cite{Walker2014}. The authors suggested a mechanism of vacancy creation based on early results on core hole decay via an Auger process \cite{Knotek1978, Knotek1979}. The mechanism involves five consecutive steps: (i) an impinging photon creates a Ti$3p$ core hole, (ii) the core hole is filled by O$2p$ electrons via an Auger process, (iii) after the removal of electrons O$^{2-}$ can change to O$^{+}$, (iv) Coulomb repulsion from neighboring Ti$^{4+}$ favors O$^{+}$ desorption, (v) an oxygen vacancy is created. As the first step of the process requires a threshold energy of 38 eV, there is no surprise that high carrier densities can be achieved only for $h\nu>38$ eV. Carrier densities can be subsequently decreased at will via exposure to a partial pressure of O$_{2}$ \cite{Walker2014}. As described in a previous section, the formation of the metallic state on irradiation is accompanied by changes in O-related spectral features. The intensity changes of O$1s$ and O$2p$ states are however disproportionate \cite{Plumb2014}. This signifies that a simple chemical doping scenario is not sufficient to describe the transition to a metallic state, as it must be accompanied by orbital or spatial changes in the near-surface region.

In the same study, the Plumb et al. revisited the dimensionality of the 2DES states \cite{Plumb2014}. They proposed that although the $d_{xy}$ states are essentially 2D, the $d_{xz}$ / $d_{yz}$ states are more 3D-like, albeit with an out-of plane dispersion that differs significantly from bulk expectations. As a result, it was concluded that $d_{xz}$ / $d_{yz}$ electrons penetrate multiple unit cells towards the bulk, while d$_{xy}$ electrons are more confined in the near-surface region. Tight binding supercell calculations by King et al. are in good agreement with this conclusion \cite{King2014}. It was shown that out-of-plane potential variations act only as a weak perturbation on electronic states that have large hopping amplitudes along the same direction. In other words, out-of-plane orbitals (i.e. $d_{xz}$ / $d_{yz}$) are less affected by such a potential and they therefore retain some 3D character confirming the aforementioned ARPES results.

The combined ARPES + tight binding study by King et al. brought into light further details of the electronic band structure of the 2DES. The energy dispersion was shown to be a result of orbital ordering, spin splitting and many body interactions \cite{King2014}. Orbital ordering refers to the energy splitting of the $d_{xy}$ and $d_{xz}$ / $d_{yz}$ states which are degenerate in the bulk. As first shown in the work by Santander et al., lift of degeneracy and orbital ordering are natural consequences of confinement along the out-of-plane direction \cite{Santander2011}. According to Eq. (1), states with different masses in the out-of-plane direction will acquire different energies on confinement. From the experimental Fermi surface areas one can derive the orbital ordering polarization $P=\frac{n(d_{xy})-n(d_{xy/yz})}{n(d_{xy})+n(d_{xy/yz})}$, where $n$ stands for the carrier density. In the case of STO(100), $P$ exceeds 30\%. Apart from orbital ordering, breaking of inversion symmetry can lift the spin degeneracy of the surface-confined states through a Rashba-Bychkov (RB) spin-orbit interaction \cite{Bychkov1984, Ast2007, Frantzeskakis2010, Frantzeskakis2010_2}. The tight-binding supercell calculations by King et al. suggest that the RB spin splitting is too weak to be observed experimentally. It reaches a maximum value of a few meV at the hybridization points of heavy (dominant $d_{xz}$/$d_{yz}$ character) and light (dominant $d_{xy}$ character) bands \cite{King2014}. Both spin angular momentum and orbital angular momentum abruptly change sign across such hybridization gaps. The possibility of a larger RB spin splitting was investigated in later studies and the main results will be reviewed in a following pragraph. On top of orbital ordering and RB spin splitting, many-body effects modify the energy dispersion of the 2DES states \cite{King2014}. Electron-phonon and electron-electron interactions renormalize the dispersion, resulting in a substantial increase of the effective mass at $E_{\textmd{F}}$. 

\begin{figure}
  \centering
  \includegraphics[width = 8.5 cm]{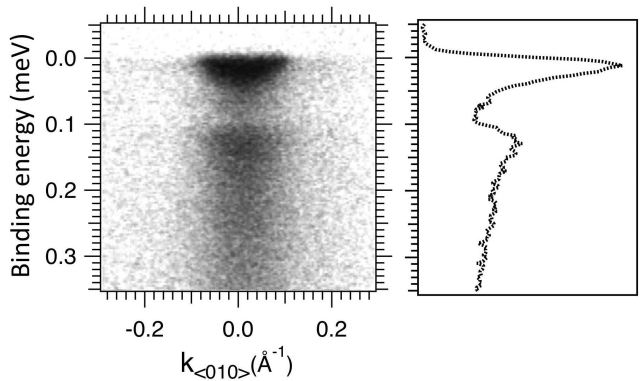}
  \caption{(left) Near-$E_{\textmd{F}}$ electronic band structure of the 2DES on SrTiO$_{3}$(100) at low carrier densities ($n_{\textmd{2D}}$ below 10$^{14}$ cm$^{-2}$). The data can be compared with the high-$n_{\textmd{2D}}$ case shown in Figs. 1(c) and 10(a). There is a replica band at an approximate binding energy of 130 meV: the fingerprint of Fr\"ohlich polarons. (right) The angle-integrated photoemission spectrum shows the characteristic peak-dip-hump lineshape. Data have been acquired using 47 eV photons and linear vertical polarization.}
\end{figure}

\begin{figure*}
  \centering
  \includegraphics[width = 13 cm]{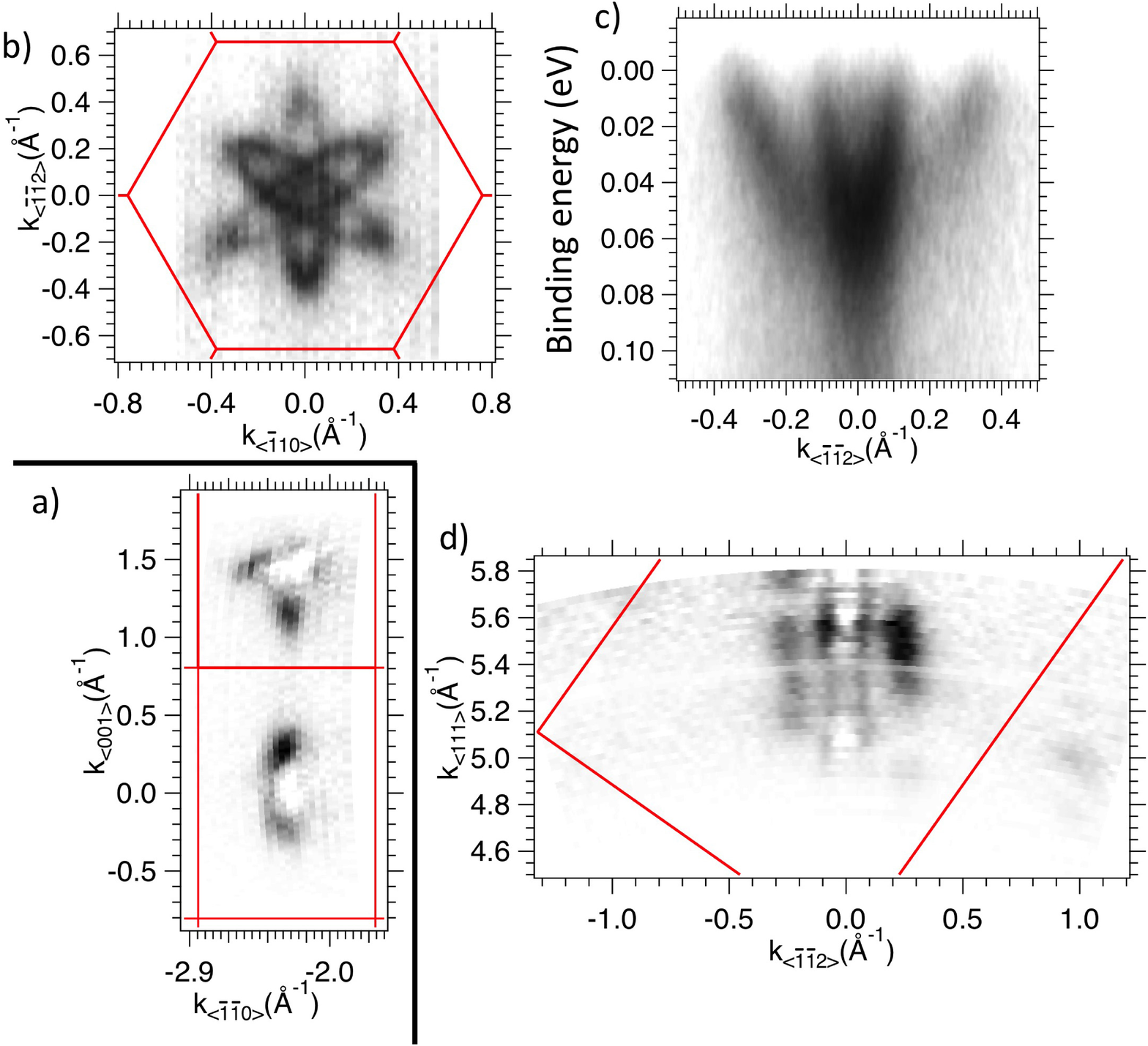}
  \caption{(a) The Fermi surface of SrTiO$_{3}$(110) in the surface plane. The contours are two orthogonal ellipses that become enhanced in adjacent Brillouin zones due to photoemission selection rules. (b) The Fermi surface of SrTiO$_{3}$(111) in the surface plane. (c) The near-$E_{\textmd{F}}$ electronic band structure of SrTiO$_{3}$(111) around the center of the Brillouin zone. (d) Fermi surface contours of SrTiO$_{3}$(111) obtained along a plane normal to the surface by varying the photon energy between 67 eV and 120 eV. States are dispersing very weakly confirming their quasi-2D character. In all cases, red lines denote the borders of the surface Brillouin zone. Data shown in panels (a) and (b)/(c) have been respectively acquired using 91 eV and 110 eV photons. Part of the figure has been adapted from Ref. \onlinecite{Rodel2014}.}
\end{figure*}

The electron-phonon interaction in the 2DES states of STO has attracted significant scientific interest already before the experimental demonstration of a 2DES on its surface. Meevasana et al. reported moderate electron-phonon coupling and stressed the absence of polarons on the (100) surface of lightly-doped STO \cite{Meevasana2010}. On the contrary, Chang et al. observed a peak-dip-hump lineshape in the near-$E_{\textmd{F}}$ electronic states, where the energy range of the hump and its temperature dependence are in good agreement with the formation of a polaron due to electron-phonon interaction \cite{Chang2010}. Later studies solved this apparent controversy by demonstrating that the nature of electron-phonon interaction in STO(100) depends heavily on the carrier density of the 2DES ($n_{\textmd{2D}}$) \cite{Wang2016, Chen2015}. Wang et al. showed that there are two main regimes. If $n_{\textmd{2D}}$ is low (i.e. low 10$^{13}$ cm$^{-2}$), evenly spaced band replicas are observed below the main bands of the 2DES. The replicas are attributed to Fr\"ohlich polarons, quasiparticles formed by an excess electron dressed by a polarization cloud. These quasiparticles extend over several lattice sites and propagate through the lattice as a free electron with an enhanced effective mass \cite{Devreese2009}. The existence of coherent polarons implies long-range coupling to a single longitudinal optical phonon branch. Figure 3 is an example of the replica bands on SrTiO$_{3}$(100) and the peak-dip-hump lineshape. On the other hand, if $n_{\textmd{2D}}$ is high (i.e. high 10$^{13}$ cm$^{-2}$ and low 10$^{14}$ cm$^{-2}$), electron-phonon coupling becomes weaker and of a different nature. Replica bands are no more observed and there is instead a kink in the energy dispersion of the 2DES states. These are spectroscopic fingerprints of the suppression of the long-range Fr\"ohlich interaction and a crossover to a short-range electron-phonon interaction \cite{Wang2016}. A related study by Chen et al. reported the existence of Fr\"ohlich polarons for carrier densities smaller than 7.6 $\times$ 10$^{13}$ cm$^{-2}$ \cite{Chen2015}. At variance with the results by Plumb et al. \cite{Plumb2014}, the authors suggest an increase of $n_{\textmd{2D}}$ with increasing annealing temperature. In such a scenario the existence -or not- of Fr\"ohlich polarons could be decided already during the preparation of a clean surface. In agreement with the study by Wang et al. \cite{Wang2016}, Chen and coworkers do not observe Fr\"ohlich polarons at high $n_{\textmd{2D}}$ \cite{Chen2015}. They attribute this fact not to a change in the nature of electron-phonon coupling, but to a 2D-to-3D transition. The key for this transition is, according to the authors, changes in the spatial distribution of oxygen vacancies at high annealing temperatures. Both studies point out the fact that the LAO/STO interface becomes superconducting for low $n_{\textmd{2D}}$, or in other words in the regime where the electron-phonon coupling at the surface of STO is strongest \cite{Wang2016, Chen2015}: an observation that gives a hint of a possible link between electron-phonon interaction and superconductivity. 

We now turn to the possibility of large RB spin splitting on the (100) surface of STO. Theoretical studies on the magnitude of spin splitting agree that the maximum value is obtained at the hybridization point of light and heavy bands \cite{King2014, Zhong2013, Khalsa2013, Kim2013}. As mentioned previously, the calculated value is of the order of few meV. This is because Ti is a light element and hence it has a very small atomic spin-orbit coupling. Nevertheless, the observation of magnetism at the LAO/STO interface \cite{Brinkman2007, Li2011, Bert2011, Salluzzo2013} plus magneto-transport data from which spin splittings as large as 10 meV are inferred \cite{Caviglia2010}, maintain the hope of a larger RB effect that goes beyond considerations of the atomic spin-orbit coupling. Indeed a collaborative spin-resolved ARPES (SARPES) study at the Swiss Light Source between CSNSM and the Paul Scherrer Insitute (PSI) concluded that the $d_{xy}$ bands of the 2DES are spin polarized and exhibit opposite chiralities, as if they were the two counterpart states in a RB model \cite{Santander2014}. The peculiar characteristics of these results are the magnitude of the experimental spin splitting and the unconventional topography of the spin-polarized subbands. The experimental data were modeled by a combined RB + Zeeman splitting where the Rashba parameter $\alpha_R$ has a value of 1 order of magnitude larger than what is expected for the calculated electric field at the surface of STO (i.e. 83 MeV/m, see subsection A). As mentioned in Ref. \onlinecite{Santander2014}, this result is surprising and -inevitably- at odds with the conclusions of previous studies concerning the origin of the near-$E_{\textmd{F}}$ electronic states: if the $d_{xy}$ subbands were due to a very large spin splitting and not because of electron confinement, the concept of the 2DES may have to be revisited. However, new experimental results \cite{Walker2016} did not reproduce these findings. McKeown Walker et al. performed SARPES experiments using the PHOENEX endstation at the BESSY II synchrotron and showed that spin polarization of the 2DES states falls within the noise level of the instrument as it is at least one order of magnitude smaller than the value proposed in Ref. \onlinecite{Santander2014}. Such large discrepancies call for further experimental efforts. We note that different protocols for the 2DES generation were followed in each experiment: in-situ prepared surfaces of undoped STO(100) at the SLS, fractured surfaces of lightly-doped STO(100) at BESSY. Measurement conditions in terms of photon energy and polarization were also different. Despite such differences in clean surface preparation and in other experimental details, the electronic band structure of the 2DES should be essentially identical following the universality suggested in earlier studies \cite{Santander2011,Plumb2014}. It is therefore surprising to have such large differences in the spin texture and more SARPES experiments are required in this direction.

In summary, the STO(100) surface offers a tunable 2DES with high carrier density, mixed dimensionality, orbital ordering, many body interactions and a potentially large spin splitting. It is therefore no surprise that experimental efforts were extended to different surfaces of STO and to other TMOs.
 \\

\subsection{2DESs on SrTiO$_{3}$ surfaces with different orientations}

\begin{figure}
  \centering
  \includegraphics[width = 6 cm]{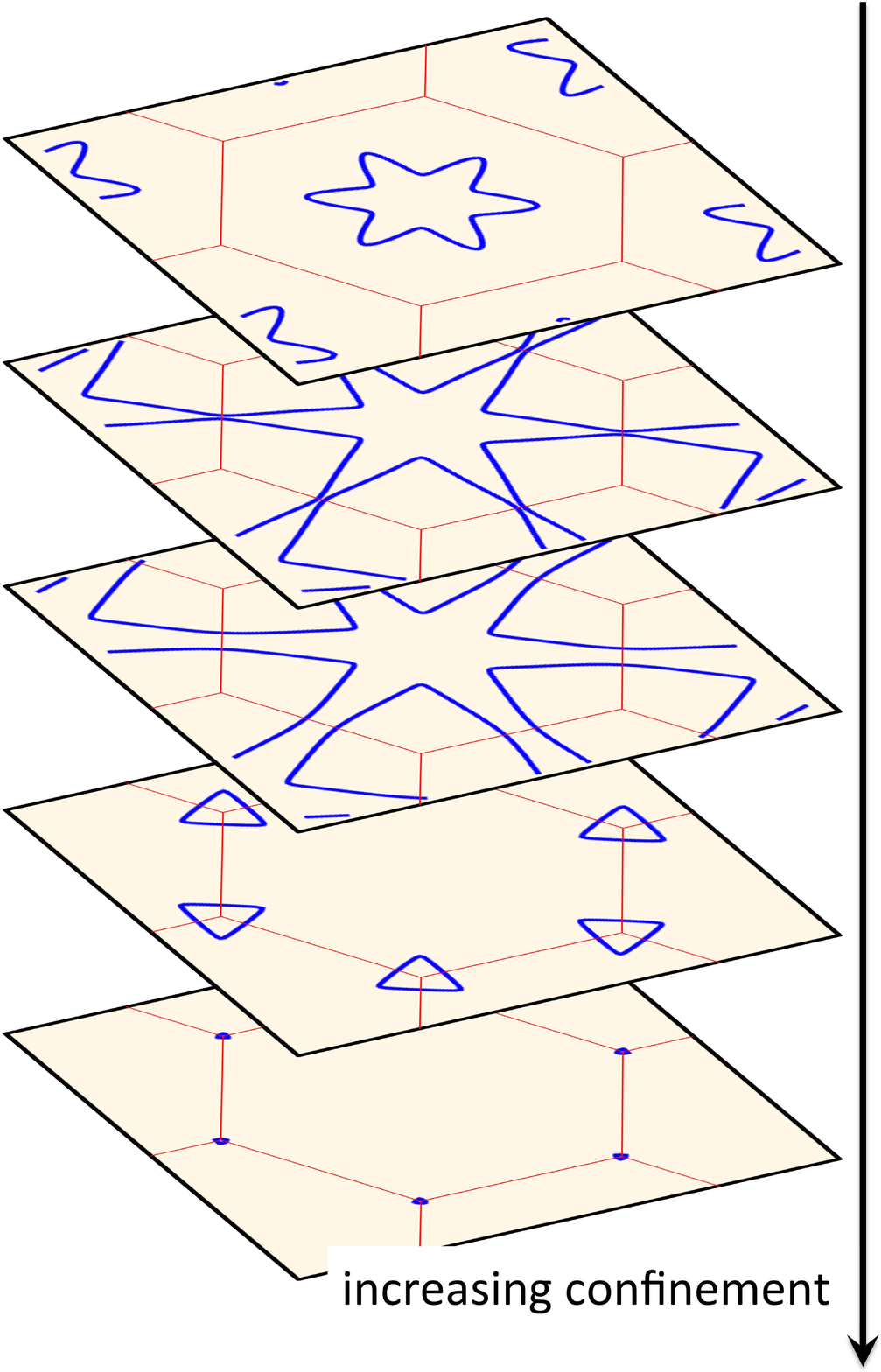}
  \caption{A cartoon based on tight-binding calculations that shows the evolution of the Fermi surface of SrTiO$_3$(111) on increasing confinement. By a rigid energy shift, the outer contour changes gradually into a graphene-like Fermi surface consisting of non-negligible intensity only at the corners of the surface Brillouin zone ($\overline{\textmd{K}}$ points). The middle panel is reminiscent of the Fermi surface of KTaO$_3$(111) (see Fig. 7).}
\end{figure}

In this section we will discuss experimental ARPES results on bare surfaces of STO having different orientations. The motivations for these studies were on one hand the successful experiments on STO(100) and on the other hand the theoretical prediction of topological states at (111) and (110) interfaces of TMOs \cite{Xiao2011, Ruegg2011, Yang2011}. The concomitant discovery of 2DESs with novel characteristics at the (111) and (110) interfaces of LAO/STO \cite{Herranz2012, Annadi2013} further fuelled these efforts. 

\begin{figure*}
  \centering
  \includegraphics[width = 13 cm]{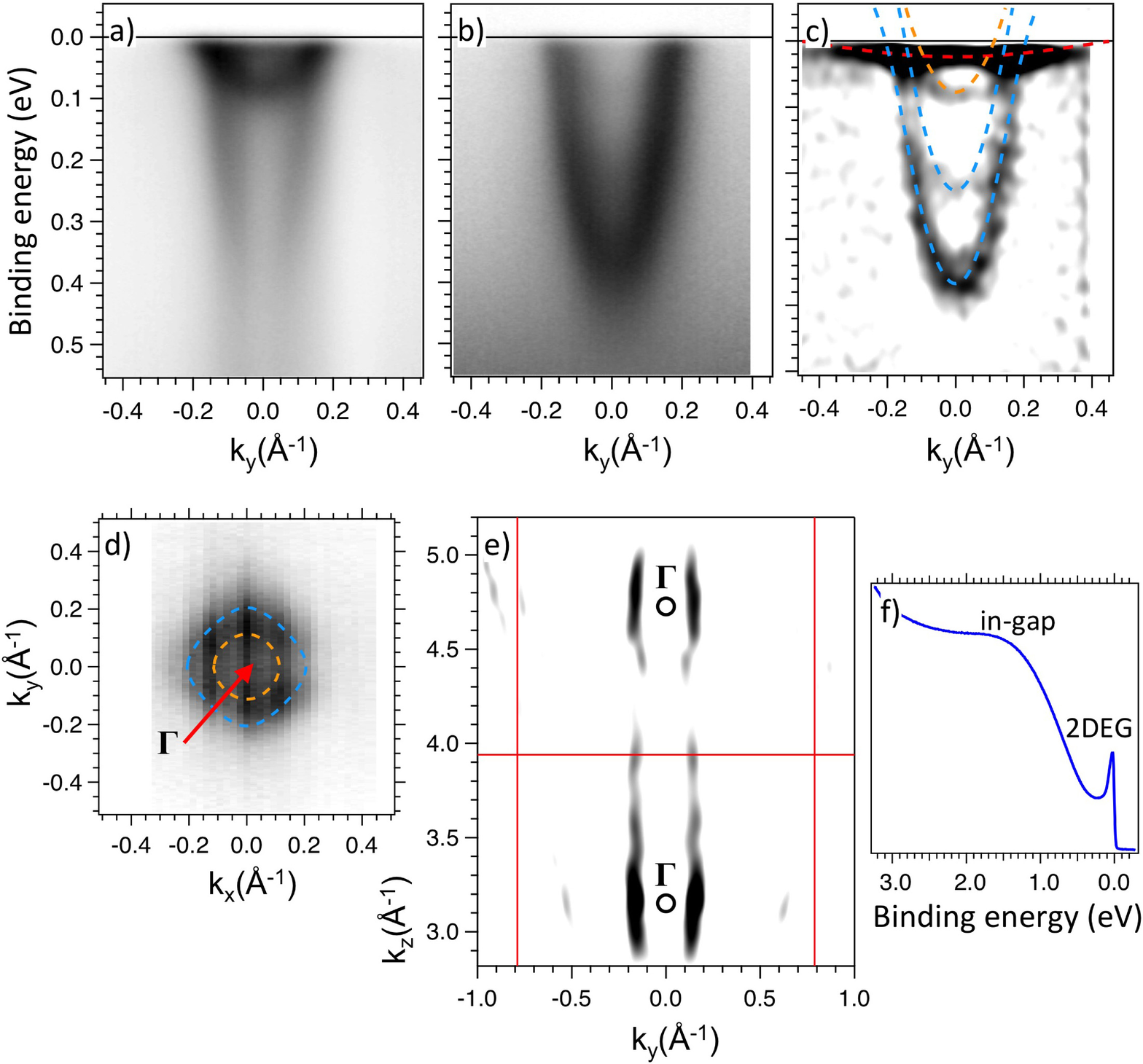}
  \caption{(a), (b) Energy-momentum ARPES intensity maps of the 2DES subbands in KTaO$_{3}$(100) acquired around the center of different surface Brillouin zones. (c) Same as (b) showing the 2$^{\textmd{nd}}$ derivative of the ARPES intensity. Dashed lines are guides to the eye and color denotes bands of different origin. (d) The Fermi surface KTaO$_{3}$(100) in the surface plane. Dashed contours are guides to the eye where the color corresponds to the bands traced in panel (c). (e) The Fermi surface of KTaO$_{3}$(100) obtained along a plane normal to the surface by varying the photon energy. States disperse very weakly confirming their two-dimensional character. (f) Angle-integrated photoemission spectrum showing that the generation of a 2DES on KTaO$_{3}$(100) is accompanied by an in-gap state, as in the case of SrTiO$_3$ surfaces (see Fig. 2). Data have been acquired with 32 eV [panels (a), (d)] and 41 eV photons [panels (b), (c)]. The polarization was linear horizontal. Part of the figure has been adapted from Ref. \onlinecite{Santander2012}.}
\end{figure*}

\begin{figure}
  \centering
  \includegraphics[width = 6.5 cm]{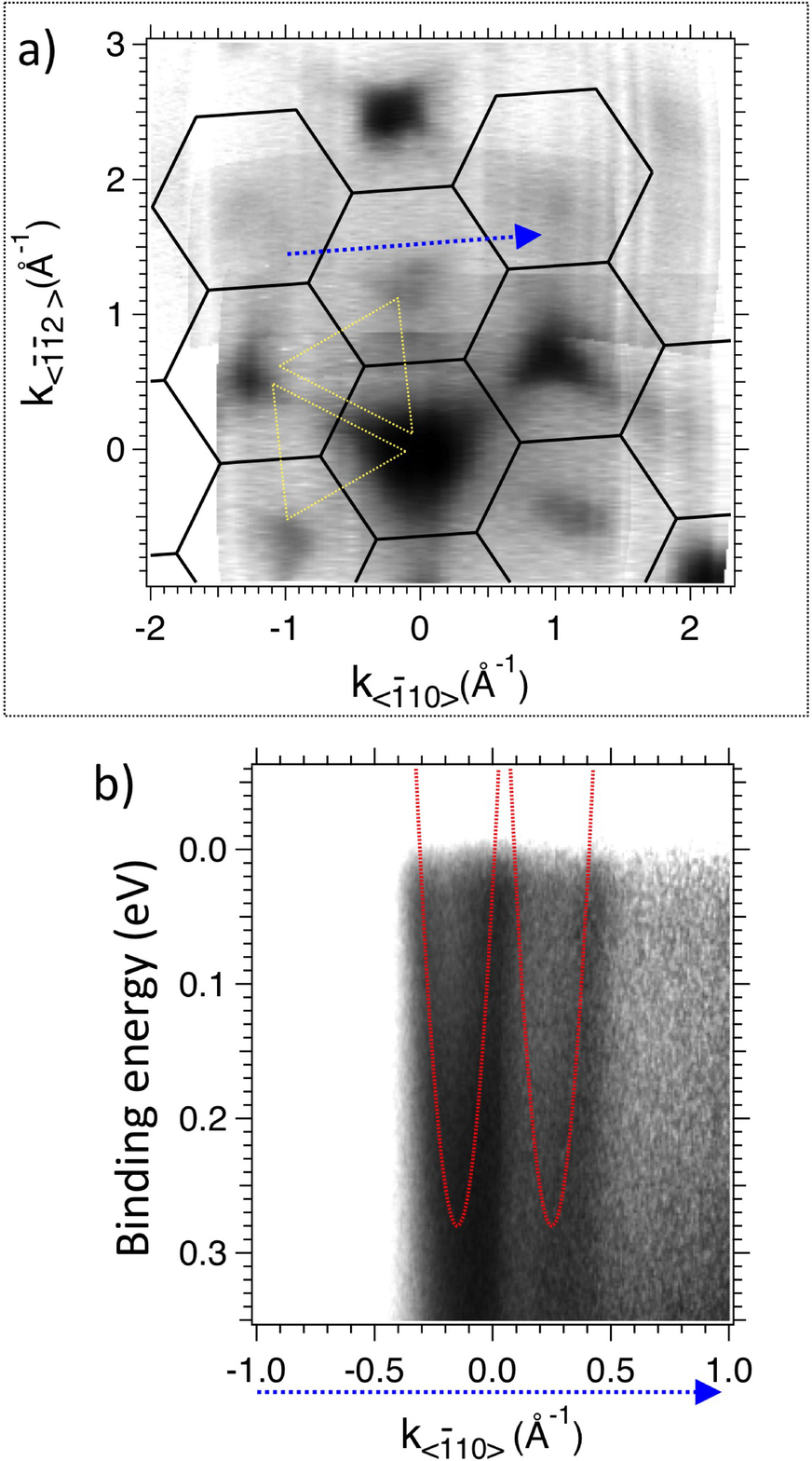}
  \caption{(a) The experimental Fermi surface of KTaO$_{3}$(111) in the surface plane. Dashed contours are approximative guide to the eyes showing the formation of an electron gutter between two contours of triangular symmetry. The results consist of a superposition from data acquired using 96 eV and 50 eV photons. A schematic version of the Fermi surface contours obtained on KTaO$_{3}$(111) is also shown in the central panel of Fig. 5. (b) Experimental electronic structure along the direction marked by a dashed blue arrow in panel (a). Red dashed contours are parabolic fits to the experimental ARPES data. Data in panel (b) have been acquired with $h\nu=$96 eV. Part of the figure has been adapted from Ref. \onlinecite{Bareille2014}.}
\end{figure}

In contrast to the STO(100) surface, the (110) surface of STO is highly polar. It therefore spontaneously reconstructs to avoid a diverging potential \cite{Nakagawa2006}. The Fermi surface has twofold symmetry and it consists of two orthogonal ellipsoids that correspond to $d_{xy}$ and $d_{xz}$/$d_{yz}$ orbitals as revealed by polarization dependent ARPES measurements \cite{Rodel2014, Wang2014} [Fig. 4(a)]. The observed states do not disperse along the out-of-plane direction, being therefore two-dimensional and comprising a 2DES at the (110) surface of STO \cite{Rodel2014, Wang2014}. There is no sign of additional periodicity in the electronic structure, which means that the 2DES is not affected by surface reconstructions. We have therefore proposed that it resides in subsurface layers \cite{Rodel2014}. This conclusion is further supported by first principles calculations that described the diffusion of oxygen vacancies after exposure of the STO(110) surface to synchrotron radiation \cite{Wang2014}. Namely, it was found that oxygen vacancies in STO(110) spontaneously migrate to subsurface layers, while they stay on the surface in the case of STO(100). Confinement along the [110] direction modifies heavily the characteristics of band dispersion. In fact, the effective masses are very different along the direction of confinement with respect to their bulk values \cite{Rodel2014} and even depend on the subband index $n$ \cite{Wang2014}. This is different to the [100] confinement where effective masses are unchanged with respect to the bulk \cite{Rodel2014, Santander2011}. The Fermi surface is highly anisotropic and the constant energy contours are composed of only $d_{xz}$ / $d_{yz}$ orbitals at higher binding energies \cite{Wang2014}. This anisotropy is most probably related to the anisotropic transport behavior observed at the (110)-oriented LAO/STO interface \cite{Annadi2013}.

The (111) surface of STO shares a lot of common features with STO(110). First of all, it is also a highly polar surface with the same nominal charge ($\pm$4$e$) as its (110) counterpart. The existence of a 2DES is supported by the absence of out-of-plane dispersion of the near-$E_{\textmd{F}}$ electronic states [Fig. 4(d)], while its insensitivity to surface roughness and reconstructions reveals that the 2DES is localized in subsurface layers \cite{Rodel2014, Walker2014_2}. Tight-binding supercell calculations verify that the corresponding 2DES wavefunctions have negligible spectral weight in the topmost layer \cite{Walker2014_2}. The Fermi surface presents a threefold symmetry [Fig. 4(b)] consisting of three -$d_{xy}$, $d_{xz}$ and $d_{yz}$- ellipsoids. In contrast to STO(100), there is no degeneracy lift of the $d_{xy}$ and $d_{xz}$/$d_{yz}$ orbitals on confinement along the $<$111$>$ direction \cite{Rodel2014, Walker2014_2} [Fig. 4(c)]. This is because of the cubic symmetry of the underlying perovskite lattice: all the $t_{2g}$ orbitals look the same (hence produce bands with identical effective mass) along the $<$111$>$ direction. Another common characteristic with the STO(110) surface is that the effective masses of the confined states are modified with respect to their bulk values. As a matter of fact, the (111) surface projection of the bulk bands is not equivalent to a (111)-oriented cut of the bulk FS, in contrast to the case of STO(100) \cite{Rodel2014, Walker2014_2}. 

The carrier density of the 2DES formed on the (110) and (111) surfaces of STO is of the order of 10$^{14}$ cm$^{-2}$ as calculated by the size of the experimental Fermi surface contours \cite{Rodel2014, Wang2014, Walker2014_2}. While the 2DES observed on STO(110) might be the key to understand the anisotropic transport behavior of corresponding interfaces \cite{Annadi2013}, the 2DES observed on STO(111) might be the key for introducing topological characteristics in transition metal oxides. The typical perovskite structure of ABO$_{3}$ is simply a buckled 2D honeycomb lattice if a single (111)-oriented bilayer of B cations is considered. Therefore, if the asymmetry in the confinement potential is neglected (case of a surface 2DES) or engineered to vanish (artificial heterostructure), a bilayer-confined (111)-oriented 2DES is equivalent to electrons confined in a honeycomb lattice. According to the cornerstone studies of Haldane and Kane \& Mele, in the presence of spin-orbit interaction, the honeycomb lattice can give rise to a finite energy gap at the Fermi level and to in-gap topological edge states \cite{Haldane1988, Kane2004}. Such a gap has never been observed in graphene due to negligible spin-orbit interaction but it may become appreciable in bilayers of (111)-oriented TMOs as proposed by several theoretical studies \cite{Xiao2011, Ruegg2011, Yang2011, Doennig2013}. At present the spatial extension of the 2DES on the STO(111) surface is larger than a single bilayer. In Ref. \onlinecite{Rodel2014}, the minimum spatial extension has been experimentally estimated at 9 layers at the Fermi level. The corresponding spatial extension of the 2DES on the STO(110) is estimated at 6 layers. Fig. 5 is a cartoon showing how increasing confinement of the 2DES on the STO(111) surface can give rise to a graphene-like Fermi surface around the $\overline{\textmd{K}}$ points of the surface Brillouin zone. Epitaxial growth of a (111)-oriented bilayer and deposition of species that can act as electron donors are possible routes to achieve bilayer confinement of the 2DES.

A comparison of the electronic structure of the three STO surfaces gives a further argument in favor of a oxygen vacancy scenario at the origin of the 2DES. Differences in the nominal polar charge are not reflected in the surface carrier density $n_{\textmd{2D}}$. Namely, 2DESs on the (110) and (111) surfaces of STO exhibit the same carrier density as the non-polar (100) surface. If a polar catastrophe scenario was at play \cite{Nakagawa2006}, one would expect a much higher carrier density for the (110) and (111) surfaces due to their high nominal polar charge (4$e$). It is therefore clear that 2DES carriers on STO surfaces do not arise as a means of compensating the polar catastrophe but they must have a different origin. Alternate exposure of the STO(111) surface on synchrotron radiation revealed the origin of the 2DES on this surface. At first, ultraviolet radiation generates the near-$E_{\textmd{F}}$ 2DES states and all accompanying spectral features comprising the O-vacancy in-gap state, the Ti$^{3+}$ shoulder and the loss of spectral weight from the O$2p$ valence band \cite{Rodel2014, Walker2014_2}. When the surface is subsequently exposed to atomic oxygen, 2DES states and all accompanying spectral features disappear \cite{Walker2014_2, Dudy2016}. These experiments are a strong proof that the origin of the 2DES on the STO(111) surface is radiation-induced oxygen vacancies.

The most important added value from studies on different STO surfaces is that they reveal the possibility of orientational tuning of the confined electronic band structure. In other words, the direction of confinement is the most important factor determining the shape of the Fermi surface, the effective masses of the 2DES states, the orbital ordering and the presence or absence of degeneracies. For instance, both Figs. 4(a) and 4(d) show the Fermi surface along a (110) plane (or equivalent). Nevertheless, the electronic band structure is very different because the direction of confinement changes. These results therefore offer the possibility to tailor the microscopic properties of confined electron systems at the surface of STO and other TMOs by changing the direction of confinement.\\

\subsection{2DESs on various transition metal oxides: case studies of KTaO$_{3}$ and TiO$_{2}$ anatase}

Another approach to tailor the properties of confined electrons at the surfaces of TMOs is to change the oxide itself. In the present section we will review 2DESs generated on clean surfaces of KTaO$_{3}$ and TiO$_{2}$ anatase after exposing them to synchrotron radiation.

\begin{figure}[!b]
  \centering
  \includegraphics[width = 8.7 cm]{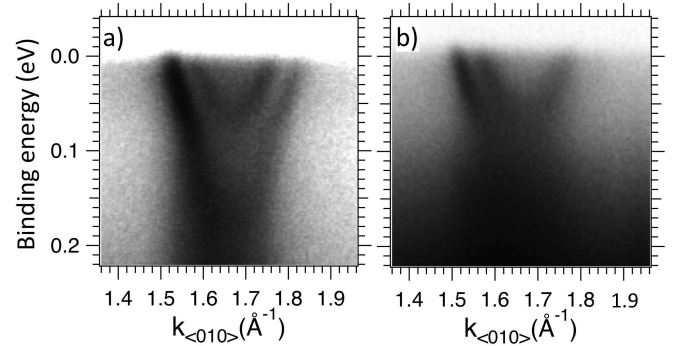}
  \caption{The experimental electronic band structure of the (100) [panel (a)] and the (110) surface [panel (b)] of TiO$_{2}$ anatase around the center of the surface Brillouin zone and along the $<$010$>$ direction. One can readily see two parabolic contours that correspond to confined subbands of $d_{xy}$ origin. The photon energy was 47 eV. More details in Ref. \onlinecite{Rodel_phd}.}
\end{figure}

KTaO$_{3}$ (KTO) is a wide gap insulator with strong spin-orbit coupling. In comparison to STO, spin-orbit coupling (SOC) in KTO is more than one order of magnitude larger because Ta is a heavier element than Ti (5$d$ vs. 3$d$). Strong spin-orbit interaction in 5$d$ TMOs can be at the origin of unconventional ground states \cite{Nakamura2009, Kim2009, Ueno2011, Pesin2010} and are of great interest in the emerging field of spintronics \cite{Zutic2004, Koo2009}. 

\begin{figure*}
  \centering
  \includegraphics[width = 15 cm]{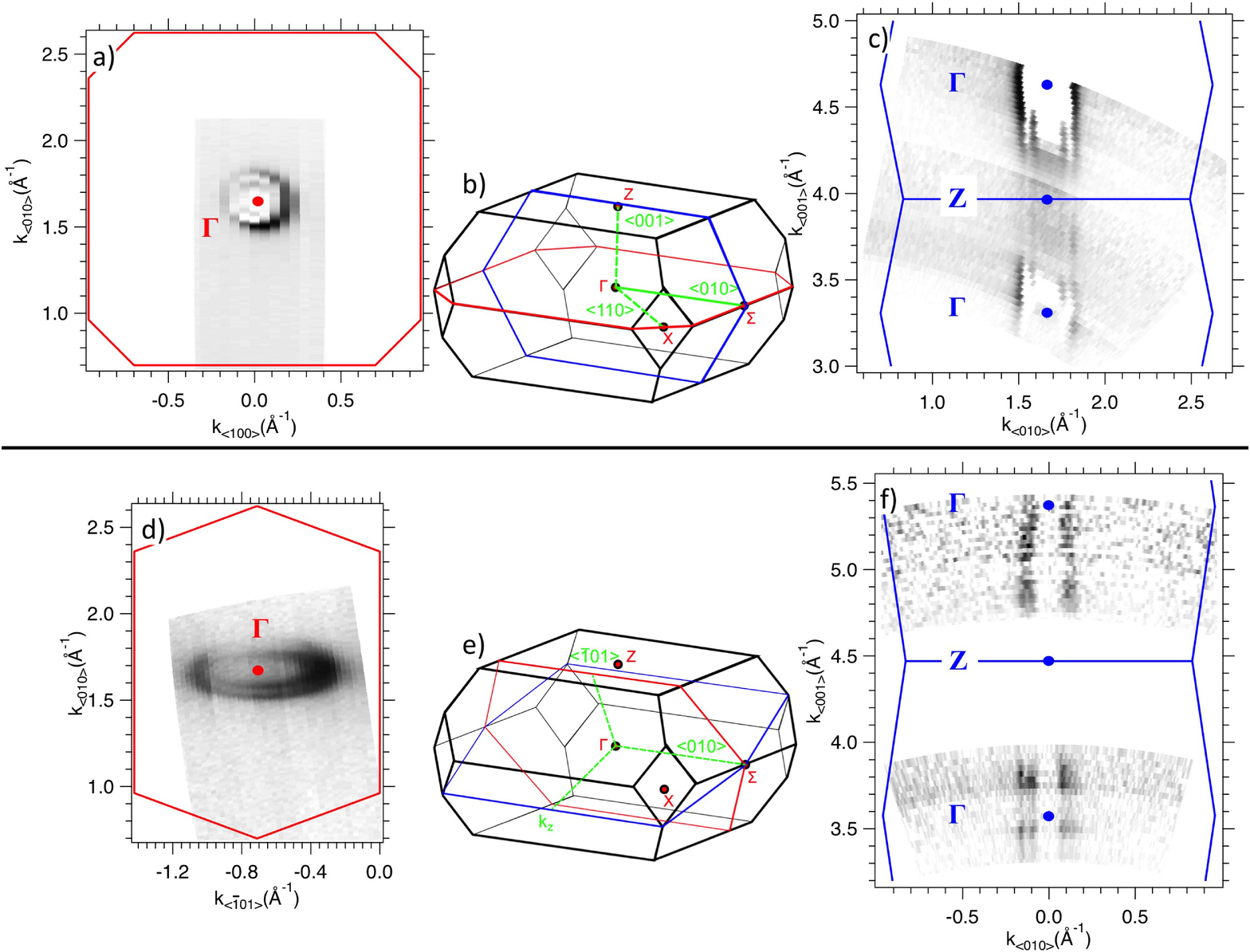}
  \caption{(a), (d) Fermi surface maps in the surface plane measured on cleaved surfaces of anatase. Panels (a) and (d) correspond to the (100) and (110) surfaces, respectively. Data is the 2$^{\textmd{nd}}$ derivative of ARPES intensity. (b), (e) The bulk Brillouin zone of TiO$_{2}$ anatase. Green arrows track the relevant high-symmetry directions, while blue and red contours show the relevant high-symmetry planes. Panels (b) and (e) correspond to the (100) and (110) surfaces, respectively. (c), (f) Fermi surface maps obtained along a plane normal to the surface of anatase by varying the photon energy. Panels (c) and (f) correspond to the (100) and (110) surfaces, respectively. Data is the 2$^{\textmd{nd}}$ derivative of ARPES intensity. The negligible out-of-plane dispersion proves the two-dimensional character of the electronic states. Experimental results in panels (a) and (d) have been acquired with 47 eV photons. More details in Ref. \onlinecite{Rodel_phd}.}
\end{figure*}

ARPES experiments on clean surfaces of KTO(100) demonstrate the existence of electronic states crossing the Fermi level [Figs. 6(a)-6(c)] \cite{Santander2012, King2012}. Those states can be well reproduced by a tight binding model of bulk bands subject to a strong spin-orbit interaction and surface confinement \cite{Santander2012}. The bands yield a closed Fermi surface in the surface plane [Fig. 6(d)] but there is no dispersion along the out-of-plane direction [Fig. 6(e)] \cite{Santander2012, King2012}. Moreover, the same near-$E_{\textmd{F}}$ states are observed for substrates with bulk doping varying by at least 5 orders of magnitude \cite{King2012}. These findings signal a 2DES at the (100) surface of KTO. Similarly to STO, the observation of an in-gap state establishes the oxygen vacancies as the origin of the 2DES \cite{Santander2012, King2012} [Fig. 6(f)]. Nevertheless, unlike the 2DES observed at the surfaces of STO, there is no evolution of the near-$E_{\textmd{F}}$ band structure on exposure to synchrotron radiation: the 2DES on KTO(100) exists right from the beginning. It is therefore possible that the polar nature of KTO(100) favors structural rearrangement that may lower the formation energy of oxygen vacancies \cite{King2012}. The effect of SOC is evident in the details of the electronic band structure. First of all, the orbital character of the $t_{2g}$ bands becomes mixed and confinement introduces an extra offset of the heavy bands with respect to the light ones \cite{Santander2012}. An additional band of very heavy mass just below $E_{\textmd{F}}$ [Fig. 6(c)] cannot be reproduced by theory. Taking into account that the cleaved surface consists of patches of (TaO$_{2}$)$^{1+}$ and (KO)$^{1-}$, this band may be attributed to electrons released by the oxygen vacancies in the KO terminations \cite{Santander2012}. Comparing the properties of the 2DESs generated on STO(100) and KTO(100), we note that they both have carrier densities of the order of low 10$^{14}$ cm$^{-2}$. Following a similar wedge model as in the case of STO(100), the confining potential on KTO(100) has a higher value at the surface ($V_\textmd{conf}(0)=-570$ meV) and the calculated electric field ($F=280$ MV/m) is almost four times stronger than on STO(100) \cite{Santander2012}. Moreover, the lower effective masses of the 2DES on KTO(100) could imply higher mobilities \cite{King2012}. The main message from the electronic band structure of KTO(100) is that mass renormalization and orbital symmetry reconstruction are possible because the values of the spin-orbit coupling, the Fermi energy and the subband splitting become comparable \cite{Santander2012}. The resulting changes of the electronic band structure have an impact on the properties of the 2DES. 

The (111) surface of KTO also exhibits a 2DES \cite{Bareille2014}. In that case, the Fermi surface has threefold symmetry and consists of a network of weakly dispersing ``electron gutters" [Fig. 7(a)]. Electron pockets come close to each other near the center of the Brillouin zone [Fig. 7(b)], while the bottom of these bands is non-dispersive along the $\overline{\Gamma \textmd{M}}$ direction. This Fermi surface can be well captured by tight binding calculations taking into account electron hopping between consecutive Ta(111) layers \cite{Bareille2014}. Its area yields a 2D carrier density $n_{\textmd{2D}} \sim 10^{14}$ cm$^{-2}$. The conduction carriers cannot be three-dimensional because in such a case the corresponding 3D carrier density $n_{\textmd{3D}} \sim 10^{21}$ cm$^{-3}$ would mean that the KTO substrate is highly-conducting and mirror-like. This is in contrast with the transparent character of the bulk crystal. From the binding energy of the confined states the upper limit of the electric field is estimated at $F>$166 MeV/m and the minimum bulk extension of the confined states at $E_{\textmd{F}}$ is 7 Ta layers \cite{Bareille2014}. We note that the observed Fermi surface resembles the simulated FS at an intermediate stage of confinement between STO(111) and a single bilayer (middle panel of Fig. 5).

As discussed in the previous paragraphs, a comparison of SrTiO$_{3}$ and KTaO$_{3}$ is instructive for the role of the transition metal atom (i.e. B) in the surface electronic properties of ABO$_{3}$ perovskite TMOs. We will now turn to the 2DES observed on the surface of a binary TMO: TiO$_{2}$ anatase. Similarly to perovskite TMOs, anatase is a transparent insulator whose crystal structure consists of oxygen octahedra around the transition metal cation (Ti$^{4+}$). There are however differences with respect to perovskites: the unit cell is body centered tetragonal, the octahedra are distorted and they are stacked by sharing their edges instead of their corners \cite{Rodel_phd}. ARPES measurements on the (100) and (101) surfaces of anatase show the presence of two bands that form tubular Fermi surfaces having closed contours in the surface plane and negligible out-of-plane dispersion (Figs. 8 and 9) \cite{Rodel2015}. This Fermi surface topography is characteristic of confined states. The observed light bands correspond to $d_{xy}$ orbitals as revealed by polarization dependent ARPES measurements \cite{Rodel2015}, in close similarity to STO. However, in contrast to the latter, there are no other bands in the occupied part of the spectrum (Fig. 8). The absence of occupied heavy bands in anatase is due to its tetragonal crystal structure that forces a lift of the bulk $t_{2g}$ degeneracy. As in STO, the 2DES in anatase is fully developed only after sufficiently long exposure to ultraviolet synchrotron radiation. The development of a 2DES is accompanied by all spectroscopic fingerprints that are typical of the creation of oxygen vacancies: in-gap vacancy-related non-dispersive state, Ti$^{3+}$ shoulder \cite{Rodel2015, Rodel_phd}. Hence, the (100) and (101) surfaces of TiO$_{2}$ anatase host 2DESs that are generated by photo-induced oxygen vacancies. The density of conduction carriers in anatase as inferred from the corresponding Fermi surface areas are $n_{2D}(100)=5.4\times10^{13}$ cm$^{-2}$ and $n_{2D}(101)=1.5\times10^{14}$ \cite{Rodel2015}. 

The energy-momentum dispersion of the $d_{xy}$ states shows signs of appreciable electron-phonon interaction \cite{Moser2013, Rodel_phd}. This interaction can be tuned with the density of the conduction carriers as it has been extensively shown in Ref. \onlinecite{Moser2013}. As in the case of STO(100) \cite{Wang2016, Chen2015}, at low densities, one observes long-range coupling and well-defined polarons. On the other hand, when the density of conduction carriers becomes higher, polarons lose coherence and dissociate into an electron system coupled to the phonons \cite{Moser2013}. We note that the study by Moser et al. reported only one $d_{xy}$ subband that formed an elongated but closed FS contour in the out-of-plane direction \cite{Moser2013}. Therefore, the authors concluded that the $d_{xy}$ states have pure bulk character in strong disagreement with the data presented in Figure 9. Discrepancies could be due to the shallow potential well in Ref. \onlinecite{Moser2013} that is unable to confine the conduction electrons. However, other unspecified experimental reasons might also affect the results. In contrast to anatase, there is no 2DES on the (110) surface of rutile even in the presence of oxygen vacancies \cite{Rodel_phd}. As rutile is another polymorph of TiO$_{2}$ consisting of oxygen octahedra with a different stacking order, we conclude that structural factors play an important role in the promotion of conduction carriers from oxygen vacancies \cite{Rodel_phd}.\\

\subsection{A new versatile technique of generating 2DESs on transition metal oxides}

\begin{figure}[!b]
  \centering
  \includegraphics[width = 8 cm]{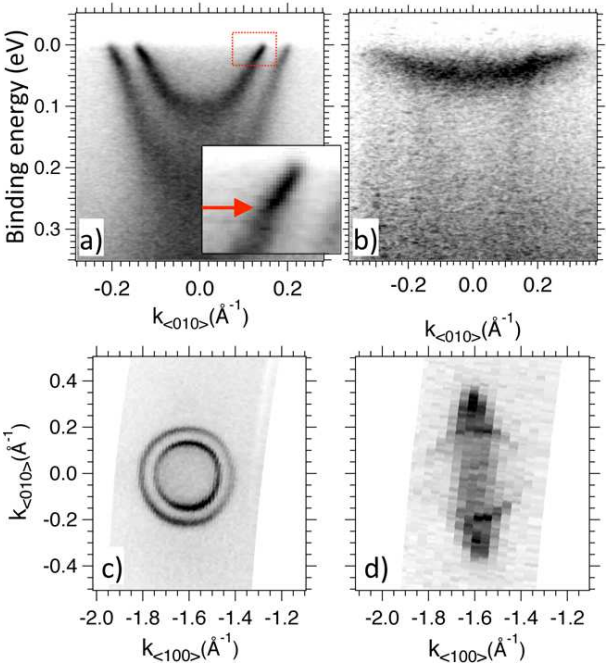}
  \caption{(a) Near-$E_{\textmd{F}}$ electronic structure of SrTiO$_{3}$(100) after the deposition of 2$\textmd{\AA}$     
  of aluminum acquired using 47 eV photons and linear vertical polarization. The inset shows the kink on the band structure due to short-range electron-phonon interaction. (b) Same as (a) using 90 eV photons and linear horizontal polarization. (c) In-plane Fermi surface of SrTiO$_{3}$(100) after the deposition of 2$\textmd{\AA}$ of aluminum acquired using 47 eV photons and linear vertical polarization. (d) Same as (c) using 90 eV photons and linear horizontal polarization. Experimental conditions in panels (a)/(c) and (b)/(d) enhance light and heavy bands, respectively. Part of the figure has been adapted from Ref. \onlinecite{Rodel2016}.}
\end{figure}

From the previous sections, it is clear that confined electron systems in TMOs possess fascinating properties that can be modified at will by changing the surface carrier density, the direction of confinement and the underlying oxide. However, with the perspective of realizing functional devices based on oxide electronics, there are important fundamental challenges that had not been tackled in experimental studies of 2DESs on TMO surfaces or interfaces. 

\begin{figure*}
  \centering
  \includegraphics[width = 13 cm]{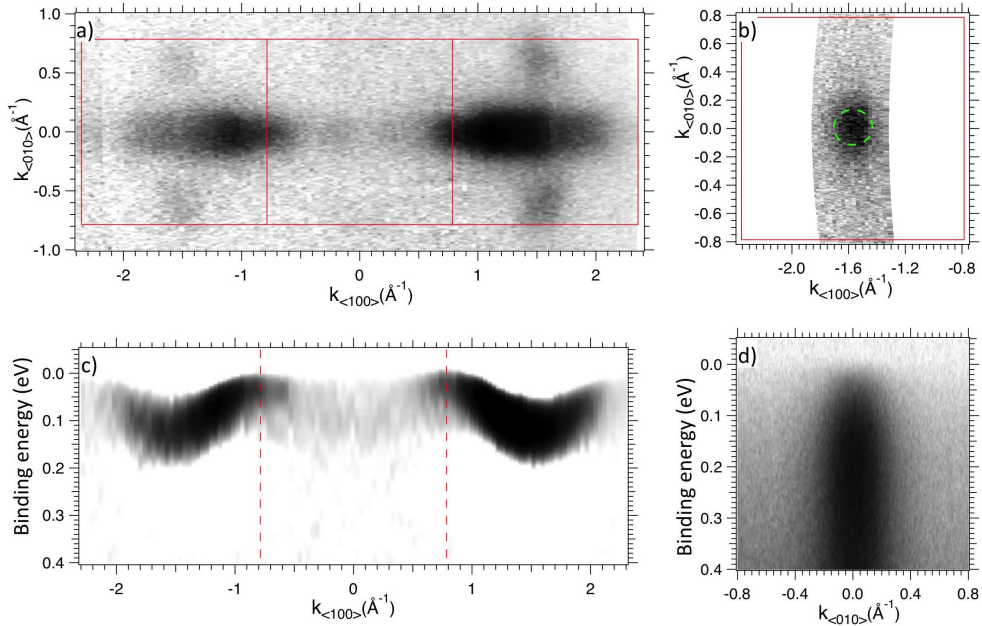}
  \caption{(a) In-plane Fermi surface of a (100)-oriented BarTiO$_{3}$ thin film after the deposition of 2$\textmd{\AA}$ of aluminum. Data is acquired using 80 eV photons and linear vertical polarization. (b) In-plane Fermi surface of a (100)-oriented BaTiO$_{3}$ thin film after the deposition of 2$\textmd{\AA}$ of aluminum. Data is acquired using 47 eV photons and linear vertical polarization. (c) Near-$E_{\textmd{F}}$ electronic structure of a (100)-oriented BarTiO$_{3}$ thin film after the deposition of 2$\textmd{\AA}$ of aluminum. Data is acquired using 80 eV photons and linear vertical polarization. (d) Near-$E_{\textmd{F}}$ electronic structure of a (100)-oriented BarTiO$_{3}$ thin film after the deposition of 2$\textmd{\AA}$ of aluminum. Data is acquired using 47 eV photons and linear vertical polarization. Experimental conditions in panels (a)/(c) and (b)/(d) enhance heavy and light bands, respectively. Part of the figure has been adapted from Ref. \onlinecite{Rodel2016}.}
\end{figure*}

Specifically, two-dimensional electron systems on interfaces require an overlayer (e.g. LaAlO$_{3}$) of thicknesses larger than 15$\textmd{\AA}$ in order to show non-zero conductivity \cite{Ohtomo2004}. Moreover, the growth procedure of the substrate/overlayer system involves deposition techniques (e.g. pulsed laser deposition) which are complex, expensive and result in properties of the 2DES that are largely dependent on all growth parameters \cite{Mannhart2010, Ohtomo2004, Chen2011}. The onset of conductivity and the growth complexity make 2DESs observed on LAO/STO and related interfaces unsuitable for mass production of oxide-based electronic devices. The discovery of 2DESs on the bare surfaces of TMOs did not answer all these challenges. Although the 2DES is not buried under an overlayer, the onset of conductivity is now due to the fact that UV irradiation is essential to create and control the 2DESs \cite{Meevasana2011, Rodel2014, Walker2014_2}. Moreover, the strong variation of the 2DES states as a function of the photon fluence received (number of photons per time per area), means that the spatial variation of the resulting 2DES depends on the size/profile of the beam spot and it can be therefore non-homogeneous. Finally, the absence of an overlayer, although favorable for the use of surface sensitive spectroscopic techniques, is an inherent disadvantage for envisaging functional devices because there is no passivation in ambient conditions. All in all, for two-dimensional electron systems in both TMO interfaces and surfaces there exist fundamental challenges that have hampered their functionalization, including an onset of conductivity (interfaces and surfaces), the complicated nature of fabrication techniques (interfaces), the lack of homogeneity (surfaces) and the absence of a passivation layer (surfaces). Such challenges were signalling the need for a different approach in fundamental experiments before the idea of functionalization might be further pursued.
 
Our team has successfully tackled all the above challenges in a very simple way. We demonstrated that 2DESs can be generated on surfaces of TMOs by room temperature deposition of aluminum \cite{Rodel2016}. Aluminum, an elementary reducing agent, pumps oxygen from the TMO substrate thereby oxidizing into insulating AlO$_{\textmd{x}}$. Most importantly, the formation of oxygen vacancies that are left behind gives rise to confined charge carriers, which are at the origin of the 2DES. The function of deposited aluminum is therefore twofold: on one hand it generates a homogeneous 2DES without the need of UV irradiation, while on the other hand it creates an oxide overlayer which acts as a passivation layer for the 2DES. As a matter of fact, having a homogeneous 2DES that requires no external trigger for its existence and is passivated with an insulating overlayer means that the aforementioned fundamental obstacles for functionalization are no longer present. On top of that, these achievements were accomplished with the thermal evaporation of pure aluminum, an approach that is straightforward, versatile and extremely cost-effective.

The ARPES results of Figure 10 show the electronic states of the 2DES generated on the (100) surface of SrTiO$_{3}$ through the deposition of Al. The confined electronic states present sharp energy-momentum dispersions near $E_{\textmd{F}}$ and high-quality Fermi surface contours. There are no spectroscopic differences with respect to the 2DESs generated by UV light proving that we have an effective -and superior- alternative to irradiation for the generation of a confined electron system. High-quality 2DESs through the deposition of Al have been successfully generated on other TMO surfaces such as SrTiO$_{3}$ (111) and anatase (100) \cite{Rodel2016}. However, the use of Al/TMO interfaces for the creation of 2DESs goes beyond a high-quality alternative to irradiation. As already mentioned, our study has shown that the 2DES is well passivated by an induced Al$_{2}$O$_{3}$ layer as long as the deposition of Al reaches a critical value \cite{Rodel2016}. This critical thickness of the deposited Al corresponds to the generation of an Al$_{2}$O$_{3}$ layer of thickness equal its natural thickness on the surface of aluminum ($\sim1.2$ nm). In this case, the 2DES electronic states are still visible in ARPES even after exposure to ambient conditions \cite{Rodel2016}. 

Another advantage of this new preparation method is the possibility to create metallic 2DESs on TMO surfaces that do not exhibit confined electron systems upon irradiation. A prime example is the (100) surface of BaTiO$_{3}$ (BTO). BTO is a well-known ferroelectric material. This ferroelectricity may coexist with metallic behavior up to a critical carrier density, when the metallic region is spatially separated from the ferroelectric one (e.g., a 2DES at the surface vs. a bulk insulating ferroelectric). Using the redox reaction of Al we succeeded in generating oxygen vacancies and hence a confined electron system on BTO(100) \cite{Rodel2016}. The 2DES/BTO system is equivalent to a metal/ferroelectric interface. Hence, polarization switching of the ferroelectric bulk might be used to control the surface conduction as in devices based on ferroelectric resistive switching \cite{Kim2013, Tra2013}. Fig. 11 shows the spectroscopic fingerprints of near-$E_{\textmd{F}}$ electronic states on the (100) surface of BTO. The Fermi surface consists of two orthogonal ellipses and a circular contour. Unlike STO(100), the elliptical contours extend beyond the borders of the surface Brillouin zone. The bands are rather broad and it is not clear whether they actually cross the Fermi level or it is the tail of non-negligible spectroscopic weight that gives rise to an apparent Fermi surface. In any case, these states have no counterpart in bulk band structure calculations and they appear only after the deposition of aluminum on the (100) surface of BTO. Details of their dispersion and dimensionality might be the subject of future studies.

The proposed new method is versatile and gives new perspectives regarding the possible combinations of reducing agents and TMO surfaces. Looking back to the Al/SrTiO$_{3}$ interface, neither the AlO$_{\textmd{x}}$ capping oxide nor the SrTiO$_{3}$ substrate possess properties (e.g. ferromagnetism, superconductivity, non-trivial topologies) that can potentially enhance the functionalities of the created 2DES/TMO/capping-oxide system. The exploration of ``active" substrates (e.g. ferroelectric BaTiO$_{3}$) and reducing agents opens new directions of research in the field of confined electron systems on oxide surfaces.\\

\section{Conclusions}

The field of confined electron systems on the surface of transition metal oxides is no longer in its infancy. The groundbreaking discovery of conducting electron systems on the bare surface of SrTiO$_{3}$(100) was soon followed by the establishment of photo-induced oxygen vacancies as the key element behind the generation of the conduction carriers. Later studies revealed that confined electron systems on transition metal oxides offer an exciting playground where orbital ordering, electron-phonon interactions and possible giant spin splitting coexist. ARPES measurements have played a central role in the development of the story, since they can give direct access to the electronic structure that arises from the confined electrons. One can therefore say that not only the 2DESs on TMO surfaces have evolved into an independent scientific field, but also the acquired ARPES knowledge can give insights on the electronic structure of the 2DESs found on TMO interfaces: a neighboring field where spectroscopic techniques with high surface sensitivity have limited hope for success. 

Such confined electron systems are not limited to the (100) surface of SrTiO$_{3}$. Different surface orientations and substrates have permitted a detailed engineering of the confined electronic structure with respect to a custom-tailored orientational tuning [(110) and (111) surfaces], strong spin-orbit interaction (KTaO$_{3}$ surfaces) and different crystal symmetries (TiO$_{2}$ anatase).

This -so far- ARPES-driven scientific field has been recently become very promising for further studies in ambient conditions with a view to device fabrication. Confined electron systems can be simply generated by the deposition of an elemental reducing agent on the surface of a transition metal oxide. The major advantage of this fabrication technique is that the generated 2DES is capped by the natural oxide of the reducing agent. Functional devices based on confined electron systems on multifunctional oxides (ferroelectric or ferromagnetic) may be therefore studied in ambient pressure due to an oxide capping layer. This simple, cost-effective and versatile technique gives rise to spatially homogeneous electron systems and overcomes previous limitations of 2DESs generated at TMO interfaces and by UV irradiation. The large opportunities opened by the numerous combinations of TMO substrates + reducing agents show the future directions of research in the field.\\

\section*{Acknowledgements}
We thank Marc Gabay, Marcelo Rozenberg, Roser Valent\'{\i}, Rub\'en Weht, C\'edric Bareille, Fran\c{c}ois Bertran and Patrick Le F\`evre for many fruitful discussions.
This work was supported by public grants from the French National Research Agency (ANR), 
project LACUNES No ANR-13-BS04-0006-01, 
and the ``Laboratoire d'Excellence Physique Atomes Lumi\`ere Mati\`ere'' 
(LabEx PALM projects ELECTROX and 2DEG2USE) overseen by the ANR as part of the 
``Investissements d'Avenir'' program (reference: ANR-10-LABX-0039).
T.~C.~R. acknowledges funding from the RTRA--Triangle de la Physique (project PEGASOS).
A.F.S.-S. thanks support from the Institut Universitaire de France.

\end{document}